# Error and jitter effect studies on the SLED for BEPCII-linac

PEI Shi-Lun(裴士伦)[1])  LI Xiao-Ping(李小平)  XIAO Ou-Zheng(肖欧正)

Institute of High Energy Physics, Chinese Academy of Sciences, Beijing 100049, China

**Abstract** RF pulse compressor is a device to convert a long RF pulse to a short one with much higher peak RF magnitude. SLED can be regarded as the earliest RF pulse compressor used in large scale linear accelerators. It is widely studied around the world and applied in the BEPC and BEPCII linac for many years. During the routine operation, the error and jitter effects will deteriorate the SLED performance either on the output electromagnetic wave amplitude or phase. The error effects mainly include the frequency drift induced by cooling water temperature variation and the frequency/$Q_0$/$\beta$ unbalances between the two energy storage cavities caused by mechanical fabrication or microwave tuning. The jitter effects refer to the PSK switching phase and time jitters. In this paper, we re-derived the generalized formulae for the conventional SLED used in the BEPCII linac. At last, the error and jitter effects on the SLED performance are investigated.

## 1. Introduction

RF pulse compressor is a device to convert a long RF pulse to a short one with much higher peak RF magnitude, with which RF pulse with much higher peak power but shorter pulse length than the klystron source developed nowadays can be obtained. SLED (SLAC Energy Doubler) is the earliest RF pulse compressor used in large scale linear accelerators [1-2]. It is widely used in the BEPC and BEPCII linac [3]. SLED uses resonant cavities with very high $Q_0$ to store energy during most duration of the incoming RF pulse. Just at one compressed-pulse-time (usually equals the filling time of the travelling wave accelerating structure) from the incident RF pulse end, the phase of the incoming pulse is reversed 180° implemented by the PSK (Phase Shift Keying) switcher. At this time, the field reflected from the cavity coupling hole then adds with, rather than subtracting from, the field emitted from the energy storage cavity constructively, and the stored energy is extracted. Due to the finite filling time of the cavity, the emitted wave cannot change instantaneously [4]; the outgoing wave experiences a sudden amplitude increase. By this way, RF power multiplication is accomplished.

Generally, SLED has two resonant cavities (Fig. 1 shows the schematic layout); they are fed by a 3dB directional coupler, so the outgoing power is directed away from the klystron [1]. One distinctive feature of the SLED is the exponential spike of the compressed pulse characterized by the resonant time constant of the cavities.

During the routine operation, the error (the frequency drift caused by cooling water temperature variation, the unbalances of the frequency, the unloaded quality factor $Q_0$ and the coupling coefficient $\beta$ between the two cavities caused by mechanical fabrication or microwave tuning) and jitter (the PSK switching phase/time jitter) effects will deteriorate the SLED performance either on the output wave amplitude or phase. On the other hand, these error and jitter effects may result in power reflection to the klystron source. In this paper, we studied the error and jitter effects theoretically. At last, the error and jitter effects on the SLED performance for BEPCII linac (Table 1 shows the technical specifications of the BEPCII SLED [5]) are analyzed. Moreover, the fabrication and tuning error of the 3-dB directional coupler will also have some effects on the SLED performance; however it is not included in this paper limited by the paper length, so a perfect 3-dB coupler is assumed in our studies.

Table 1: Technical specifications of the BEPCII SLED

| Parameters | Values | Unit |
|---|---|---|
| Frequency | 2856 | MHz |
| $Q_0$ | 100000 | |
| $\beta$ | 5 | |
| Frequency tuning range | ±500 | kHz |
| Operating temperature | 45±0.2 | °C |
| RF pulse length | 4 | μs |
| PSK switching time | 3 | μs |
| PSK switching speed | <50 | ns |
| Compressed pulse length | 1 | μs |

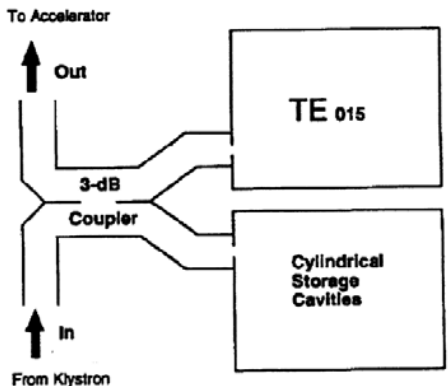

Fig. 1: The SLED schematic layout

1) E-mail: peisl@mail.ihep.ac.cn

## 2. Theory

### 2.1 SLED

By the law of power conservation, it can be given that

$$P_{in} = P_{out} + P_c + dU_c/dt \quad (1)$$

where $P_{in}$ is the input power from the generator (klystron source), $P_{out}$ is the outgoing power delivered to the accelerating section, $P_c$ is the power dissipated on the cavity walls, and $U_c$ is the electromagnetic energy stored in the resonant cavity.

Assuming $m$ is the proportionality coefficient between the square of the electric field and power in the external waveguide, then $P_{in}$ and $P_{out}$ can be written as

$$P_{in} = m|\vec{E}_{in}|^2 \quad (2)$$

$$P_{out} = m|\vec{E}_e + \vec{E}_r|^2 \quad (3)$$

$\vec{E}_{in} = \tilde{E}_{in}e^{j(\omega t-kz)}$ is the incident wave from the klystron; $\vec{E}_e = \tilde{E}_e e^{j(\omega t+kz+\varphi)}$ is the wave emitted from the cavity, where $\varphi = -\tan^{-1}(2Q_L(\omega/\omega_0 - \omega_0/\omega))$ is the cavity field phase shift relative to its value when driven on resonance defined by the loaded quality factor $Q_L = Q_0/(1+\beta)$ [4]; $\vec{E}_r = \tilde{E}_r e^{j(\omega t+kz)} = \Gamma \tilde{E}_{in} e^{j(\omega t+kz)}$ is the wave reflected from the waveguide-cavity coupling interface (aperature), where $\Gamma$ is the reflection coefficient. $\tilde{E}_{in}$, $\tilde{E}_e$ and $\tilde{E}_r$ are all complex numbers. Here $Q_L > Q_0 \gg 1$, $\beta \gg 1$ and small deviation of $\omega$ from $\omega_0$ are assumed.

From the definition of cavity coupling coefficient $\beta$ as the ratio of $Q_0$ to the external quality factor $Q_e$,

$$\beta = Q_0/Q_e = (\omega U_c/P_c)/(\omega U_c/P_e) = P_e/P_c \quad (4)$$

one can write $P_c$ as

$$P_c = P_e/\beta = m|\vec{E}_e|^2/\beta \quad (5)$$

Then, from the definition of $Q_0$, $U_c$ can be written as

$$U_c = \frac{Q_0}{\omega}P_c = \frac{Q_0}{\omega}\frac{m|\vec{E}_e|^2}{\beta} \quad (6)$$

with $\omega$ equal to the drive frequency, so

$$dU_c/dt = \frac{2mQ_0}{\omega\beta}|\vec{E}_e|\frac{d|\vec{E}_e|}{dt} \quad (7)$$

Suppose at time $t=0$, the incident wave reaches the resonant cavity, we have $U_c=0$ since the cavity cannot be filled instantaneously [4]. In the meantime, $\vec{E}_e$ is also zero. Then with Eqs. (1)–(3), (5) and (7), one can find that $|\Gamma|=1$. Considering there is a 180° phase shift associated with a reflection, $\Gamma=-1$. Strictly speaking, $|\Gamma|$ is slightly less than 1 because a small fraction of incident energy is transmitted and used to charge the resonant cavity.

Eq. (1) can now be modified to be

$$|\tilde{E}_{in}|^2 = |\tilde{E}_e e^{j\varphi} - \tilde{E}_{in}|^2 + \frac{|\tilde{E}_e|^2}{\beta} + \frac{Q_0}{\omega\beta}\tilde{E}_e\frac{d\tilde{E}_e^*}{dt} + \frac{Q_0}{\omega\beta}\tilde{E}_e^*\frac{d\tilde{E}_e}{dt} \quad (8)$$

Complex numbers $\tilde{E}_{in}$ and $\tilde{E}_e$ can be re-expressed in polar/exponential form $\tilde{E}_{in} = E_{in}e^{j\theta}$ and $\tilde{E}_e = E_e e^{j\psi}$. Both $E_{in}$ and $E_e$ are real numbers. Finally, Eq. (8) can be simplified as

$$\frac{\omega_0 T_c}{\omega}\frac{dE_e}{dt} + E_e = \alpha E_{in}\cos(\psi + \varphi - \theta) \quad (9)$$

with $\alpha = 2\beta/(1+\beta)$. $T_c$ is the loaded cavity time constant, which is given by $T_c = 2Q_L/\omega_0$.

During the routine operation, the incident wave to the SLED has constant amplitude just before the PSK phase reverse at one compressed-pulse-time from the RF pulse end. Then, $\theta$ can be set to be 0 at $t=0$ for simplicity. In the meantime, one has $\psi=\theta=0$. After the PSK phase switch, as the emitted wave cannot change rapidly when the incident wave phase $\theta$ is altered suddenly, $\psi$ will vary slowly, and then $\tilde{E}_e$ can be obtained by complex addition of the emitted waves resulting from different sources. With unit field used, $\tilde{E}_{in}$ can be written as follows.

$$\tilde{E}_{in} = \begin{cases} 1 & 0 \leq t \leq t_1 \\ e^{j\xi} & t_1 \leq t \leq t_2 \\ 0 & t_2 \leq t \end{cases} \quad (10)$$

$\xi$ is the shifted phase of $\theta$. Using Eqs. (9)–(10) and the condition $\tilde{E}_e = 0$ at $t=0$, $\tilde{E}_e$ can be obtained to be

$$\tilde{E}_e = \begin{cases} \alpha\cos\varphi\left(1-e^{-\omega t/\omega_0 T_c}\right) & 0 \leq t \leq t_1 \\ \left\{\alpha\cos\varphi\left(1-e^{-\omega t_1/\omega_0 T_c}\right)e^{-\omega(t-t_1)/\omega_0 T_c} + \alpha\cos(\psi(t)+\varphi-\xi)\left(1-e^{-\omega(t-t_1)/\omega_0 T_c}\right)\right\} \times \\ e^{j\psi(t)} & t_1 \leq t \leq t_2 \\ \left\{\alpha\cos\varphi\left(1-e^{-\omega t_1/\omega_0 T_c}\right)e^{-\omega(t_2-t_1)/\omega_0 T_c} + \alpha\cos(\psi(t_2)+\varphi-\xi)\left(1-e^{-\omega(t_2-t_1)/\omega_0 T_c}\right)\right\} \times \\ e^{-\omega(t-t_2)/\omega_0 T_c}e^{j\psi(t_2)} & t_2 \leq t \end{cases} \quad (11)$$

For time duration $t_1 \leq t \leq t_2$, the emitted wave is really the sum of two waves; one is the wave radiated by the energy stored in the cavity during $0 \leq t \leq t_1$, another is the one resulting from the phase reversed incident wave, which attempts to charge up the cavity to the opposite voltage. Alternatively, $\tilde{E}_e$ can be rewritten as

$$\tilde{E}_e = \begin{cases} \alpha\cos\varphi\left(1-e^{-\omega t/\omega_0 T_c}\right) & 0 \leq t \leq t_1 \\ \alpha\cos\varphi\left\{\left(1-e^{-\omega t_1/\omega_0 T_c}\right)e^{-\omega(t-t_1)/\omega_0 T_c} + \left(1-e^{-\omega(t-t_1)/\omega_0 T_c}\right)e^{j\xi}\right\} & t_1 \leq t \leq t_2 \\ \alpha\cos\varphi\left\{\left(1-e^{-\omega t_1/\omega_0 T_c}\right)e^{-\omega(t_2-t_1)/\omega_0 T_c} + \left(1-e^{-\omega(t_2-t_1)/\omega_0 T_c}\right)e^{j\xi}\right\}e^{-\omega(t-t_2)/\omega_0 T_c} & t_2 \leq t \end{cases} \quad (12)$$

Comparing Eqs. (11) and (12), $\psi(t)$ can be obtained to be

$$\tan\psi(t) = \frac{\left(1-e^{-\omega(t-t_1)/\omega_0 T_c}\right)\sin\xi}{\left(1-e^{-\omega t_1/\omega_0 T_c}\right)e^{-\omega(t-t_1)/\omega_0 T_c}+\left(1-e^{-\omega(t-t_1)/\omega_0 T_c}\right)\cos\xi} \quad (13)$$

Since $\vec{E}_{out} = \vec{E}_e + \vec{E}_r = \tilde{E}_{out} e^{j(\omega t + kz)}$, finally the normalized outgoing wave can be given by

$$\tilde{E}_{out} = \begin{cases} \alpha\cos\varphi\left(1-e^{-\omega t/\omega_0 T_c}\right)e^{j\varphi} - 1 & 0 \le t \le t_1 \\ \left\{\alpha\cos\varphi\left(1-e^{-\omega t_1/\omega_0 T_c}\right)e^{-\omega(t-t_1)/\omega_0 T_c} + \alpha\cos(\psi(t)+\varphi-\xi)\left(1-e^{-\omega(t-t_1)/\omega_0 T_c}\right)\right\} \times \\ e^{j\psi(t)}e^{j\varphi} - e^{j\xi} & t_1 \le t \le t_2 \\ \left\{\alpha\cos\varphi\left(1-e^{-\omega t_1/\omega_0 T_c}\right)e^{-\omega(t_2-t_1)/\omega_0 T_c} + \alpha\cos(\psi(t_2)+\varphi-\xi)\left(1-e^{-\omega(t_2-t_1)/\omega_0 T_c}\right)\right\} \times \\ e^{-\omega(t-t_2)/\omega_0 T_c} e^{j\psi(t_2)} e^{j\varphi} & t_2 \le t \end{cases} \quad (14)$$

or

$$\tilde{E}_{out} = \begin{cases} \alpha\cos\varphi\left(1-e^{-\omega t/\omega_0 T_c}\right)e^{j\varphi} - 1 & 0 \le t \le t_1 \\ \alpha\cos\varphi\left\{\left(1-e^{-\omega t_1/\omega_0 T_c}\right)e^{-\omega(t-t_1)/\omega_0 T_c} + \left(1-e^{-\omega(t-t_1)/\omega_0 T_c}\right)e^{j\xi}\right\}e^{j\varphi} - e^{j\xi} & t_1 \le t \le t_2 \\ \alpha\cos\varphi\left\{\left(1-e^{-\omega t_1/\omega_0 T_c}\right)e^{-\omega(t_2-t_1)/\omega_0 T_c} + \left(1-e^{-\omega(t_2-t_1)/\omega_0 T_c}\right)e^{j\xi}\right\}e^{-\omega(t-t_2)/\omega_0 T_c}e^{j\varphi} & t_2 \le t \end{cases} \quad (15)$$

## 2.2 3dB directional coupler

The scattering matrix of a perfect 3dB coupler is [1]

$$\hat{S} = \frac{1}{\sqrt{2}}\begin{pmatrix} 0 & 1 & 0 & i \\ 1 & 0 & i & 0 \\ 0 & i & 0 & 1 \\ i & 0 & 1 & 0 \end{pmatrix} \quad (16)$$

Suppose unit field from the klystron is fed into Port 1 of a SLED, then the output is

$$\frac{1}{\sqrt{2}}\begin{pmatrix} 0 & 1 & 0 & i \\ 1 & 0 & i & 0 \\ 0 & i & 0 & 1 \\ i & 0 & 1 & 0 \end{pmatrix}\begin{pmatrix} 1 \\ 0 \\ 0 \\ 0 \end{pmatrix} = \frac{1}{\sqrt{2}}\begin{pmatrix} 0 \\ 1 \\ 0 \\ i \end{pmatrix} \quad (17)$$

Now the cavities located at the two output ports will be charged up. The reflected waves $\tilde{E}_{out1}/\sqrt{2}$ and $\tilde{E}_{out2}/\sqrt{2}$ from the two cavities become new inputs for Port 2 and 4. $\tilde{E}_{out1}$ and $\tilde{E}_{out2}$ can be obtained by substituting the appropriate cavity characteristic parameters into Eqs. (14) or (15). To be simple, the reflected wave in this section means the sum of the reflected and emitted waves mentioned in Section 2.1. Finally the output can be given as follows.

$$\frac{1}{\sqrt{2}}\begin{pmatrix} 0 & 1 & 0 & j \\ 1 & 0 & j & 0 \\ 0 & j & 0 & 1 \\ j & 0 & 1 & 0 \end{pmatrix}\frac{1}{\sqrt{2}}\begin{pmatrix} 0 \\ \tilde{E}_{out1} \\ 0 \\ j\tilde{E}_{out2} \end{pmatrix} = \frac{1}{2}\begin{pmatrix} \tilde{E}_{out1} - \tilde{E}_{out2} \\ 0 \\ j(\tilde{E}_{out1} + \tilde{E}_{out2}) \\ 0 \end{pmatrix} \quad (18)$$

It can be seen that if the two cavities are completely the same, the reflected waves would cancel at Port 1 and combine at Port 3. The power is successfully directed away from the klystron source and delivered to the accelerating section.

## 3. Jitter effects

During the routine operation of SLED, jitter effects such as PSK switching phase/time jitter do exist.

### 3.1 PSK switching phase jitter

Figs. 2 and 3 show the amplitude and phase deviations of the outgoing wave $\tilde{E}_{out}$ for the BEPCII SLED during the compressed-pulse-time when the PSK switching phase is deviated from 180°. It can be seen that the field multiplication factor decreases a little bit at $t=t_1+$ but increases at $t=t_2-$. If the PSK switching phase jitter is controlled to be less than 2°, the outgoing wave phase jitter would be within ±1°.

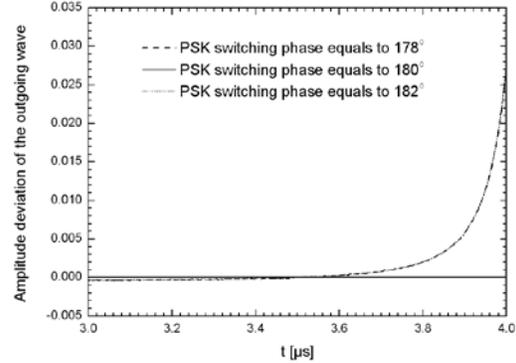

Fig. 2: Amplitude deviation of the outgoing wave $\tilde{E}_{out}$

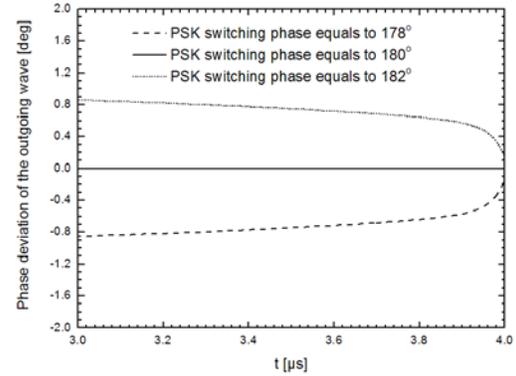

Fig. 3: Phase deviation of the outgoing wave $\tilde{E}_{out}$

### 3.2 PSK switching time jitter

PSK switching time jitter only affects the compressed-rf-pulse starting time and the power multiplication factor. Fig. 4 shows the shape of $|\tilde{E}_{out}|$ for the BEPCII SLED when the PSK switching time is varied. It clearly shows that for the same incident RF pulse, the closer the PSK switching time to the pulse end, the higher the maximum power multiplication, but the shorter the compressed-pulse-length.

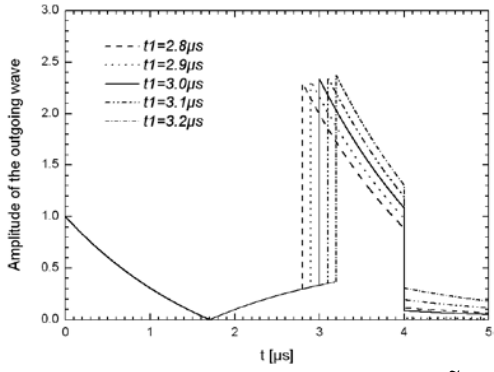

Fig. 4: Amplitude of the outgoing wave $\tilde{E}_{out}$

## 4. Error effects

In reality, the cooling water temperature must have some tolerance around the given value, and the two cavities of each SLED cannot be completely the same due to mechanical fabrication or microwave tuning error.

### 4.1 Cooling water temperature drift

For the BEPCII SLED, the cooling water temperature is controlled to be 45±0.2°C, then with the scaling law Δf/ΔT=45kHz/°C [6], the induced cavity frequency deviation from resonance is ≤±9kHz. Given the two cavities still have the same characteristics, Figs. 5 and 6 show the amplitude and phase of the outgoing wave $\tilde{E}_{out}$ when the cooling water temperature is drifted away from 45°C, the power multiplication factor decreases.

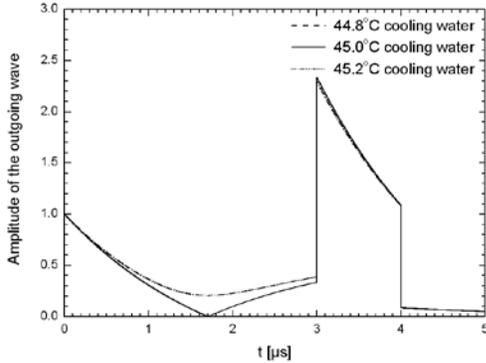

Fig. 5: Amplitude of the outgoing wave $\tilde{E}_{out}$

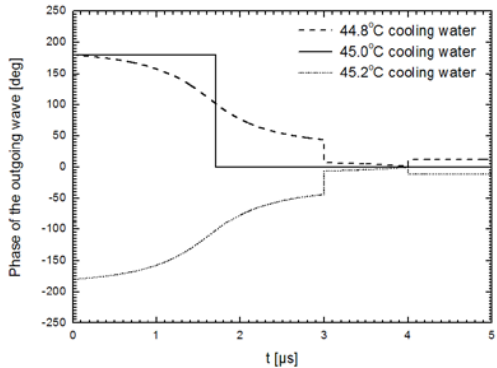

Fig. 6: Phase of the outgoing wave $\tilde{E}_{out}$

Figs. 7 and 8 show the amplitude and phase deviations of the outgoing wave during the compressed-pulse-time. From Figs. 5 and 7, it can be found that for temperature tolerance of ±0.2°C the maximum field multiplication factor decrease is ~1.67%, while the averaged one is ~0.88%. Correspondingly, the maximum and average phase deviations are ~6.8° and ~3.8° respectively.

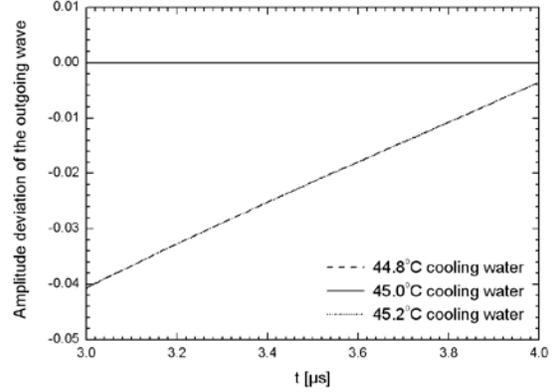

Fig. 7: Amplitude deviation of the outgoing wave $\tilde{E}_{out}$

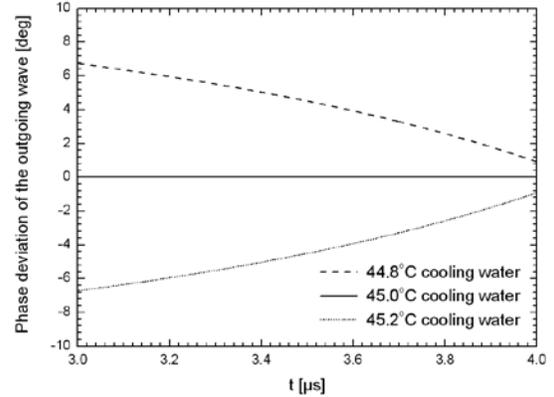

Fig. 8: Phase deviation of the outgoing wave $\tilde{E}_{out}$

### 4.2 Cavity characteristic unbalance errors

Even if the fabrication and tuning processes are strictly controlled, the cavity characteristics such as frequency, $Q_0$ and $\beta$ unbalance errors may be produced. The unbalance errors will result in differences between the two cavities' outputs $\tilde{E}_{out1}$ and $\tilde{E}_{out2}$.

#### 4.2.1 Frequency unbalance error

Figs. 9 and 10 show the amplitude and phase deviations of the outgoing wave when the frequencies of the two cavities have some small differences. The larger $|\delta f1|+|\delta f2|$, the bigger the amplitude deviation. If $|\delta f1|=|\delta f2|$ and one compares Figs. 7 and 9, the outgoing wave amplitude deviation of $\delta f1=-\delta f2$ is bigger than that of $\delta f1=\delta f2$, and the phase deviation is roughly proportional to $|\delta f1+\delta f2|$. There would not be any phase deviation if $\delta f1=-\delta f2$.

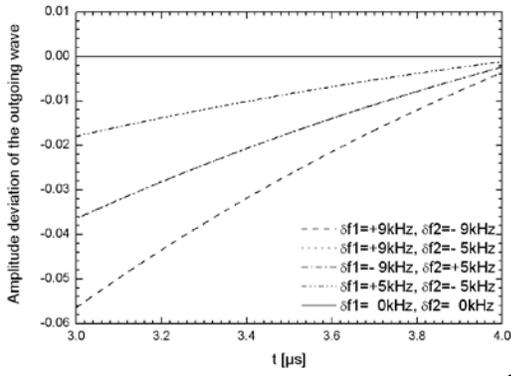

Fig. 9: Amplitude deviation of the outgoing wave $\tilde{E}_{out}$

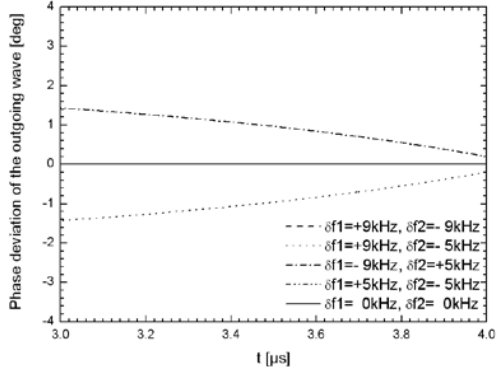

Fig. 10: Phase deviation of the outgoing wave $\tilde{E}_{out}$

Fig. 11 shows the amplitude of the reflected wave to the klystron during the whole RF pulse duration. Once there is frequency unbalance between the two cavities, there would be some power reflection to the klystron source, the reflected wave amplitude will be directly proportional to $|\delta f1 - \delta f2|$.

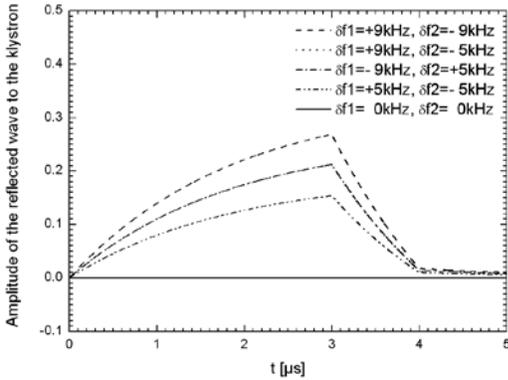

Fig. 11: Amplitude of the reflected wave to the klystron

During the microwave tuning process, the two cavities are usually tuned to resonance one by one. Fig. 12 shows the amplitudes of the outgoing wave to the accelerating structure and the reflected wave to the klystron source when only one cavity is tuned to resonance. It can be seen that part of the RF power is reflected back to klystron source, and the field multiplication factor of the outgoing wave decreases; however, the maximum outgoing wave's amplitude is still higher than half of the amplitude when both cavities are on resonance, this is because the resonance of one cavity only turns off the corresponding emitted wave, the reflected wave from that cavity still functions.

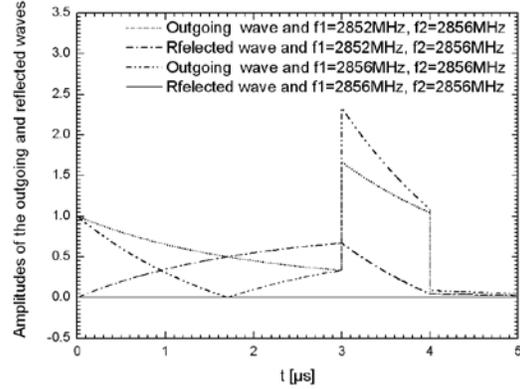

Fig. 12: Amplitudes of the outgoing and reflected waves

### 4.2.2 $Q_0$ unbalance error

For the given cavity geometry and material, $Q_0$ is mostly determined by the surface processing technique. Lower surface roughness usually gives higher $Q_0$. Fig. 13 shows the amplitude deviations of the outgoing wave when the cavities' $Q_0$s of the BEPCII SLED are deviated from the designed value of 100000. The larger the value of $|\delta Q_{01} + \delta Q_{02}|$, the bigger the amplitude deviation. If $\delta Q_{01}+\delta Q_{02}>0$, the field multiplication factor decreases at $t=t1+$ but increases at $t=t2-$. Correspondingly, if $\delta Q_{01}+\delta Q_{02}<0$, it goes the reverse way. $Q_0$ unbalance has no effect on the phase deviation except it is accompanied by the frequency unbalance. For that case, it will magnify or minify the phase deviation caused by frequency unbalance, which depends on the $Q_0$ deviation value. Higher $Q_0$ will magnify the phase deviation, and lower $Q_0$ will minify it.

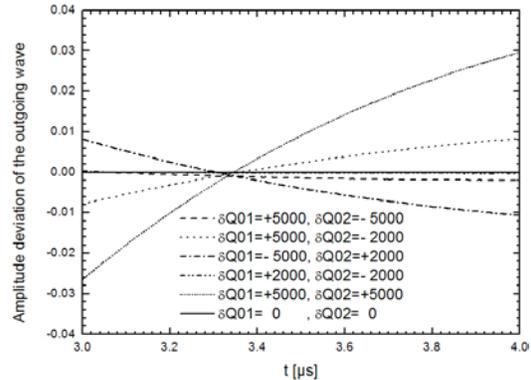

Fig.13: Amplitude deviation of the outgoing wave $\tilde{E}_{out}$

Fig. 14 shows the amplitude of the reflected wave to the klystron when $Q_0$ unbalance exists. The bigger $|\delta Q_{01} - \delta Q_{02}|$, the larger the reflection. For two cases with the same $|\delta Q_{01} - \delta Q_{02}|$, the reflection of $\delta Q_{01}+\delta Q_{02}<0$ is slightly larger than that of $\delta Q_{01}+\delta Q_{02}>0$.

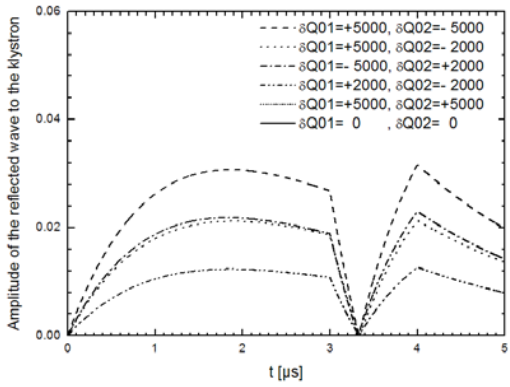

Fig. 14: Amplitude of the reflected wave to the klystron

### 4.2.3  β unbalance error

β unbalance is mainly produced by machining the cavity-waveguide coupling hole. Fig. 15 shows the outgoing wave amplitude when the two cavities have different β. Higher value of β1+β2 would give higher field enhancement at t=t1+ but lower one at t=t2-, whereas for two cases with equal β1+β2, the one with β1 ≠β2 always has lower field enhancement during the compressed pulse time than that one with β1=β2. Similar to the $Q_0$ unbalance, β unbalance also has no effect on the phase deviation except it is accompanied by frequency unbalance. Opposite to $Q_0$, higher β will minify the phase deviation, lower β will magnify the phase deviation.

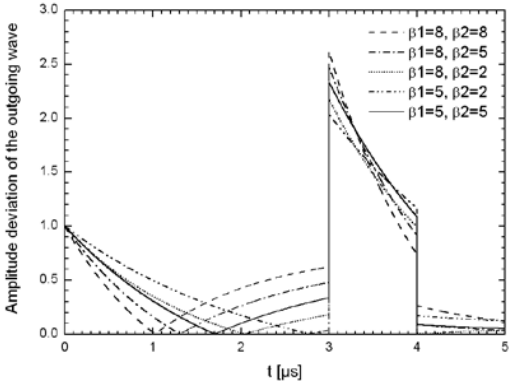

Fig.15: Amplitude of the outgoing wave $\tilde{E}_{out}$

Fig. 16 shows the reflected wave amplitude to the klystron source. Bigger $|\beta 1-\beta 2|$ gives higher reflection. For the same $|\beta 1-\beta 2|$, higher β gives lower reflection due to the smaller $|\alpha 1-\alpha 2|=|2\beta 1/(1+\beta 1)-2\beta 2/(1+\beta 2)|$.

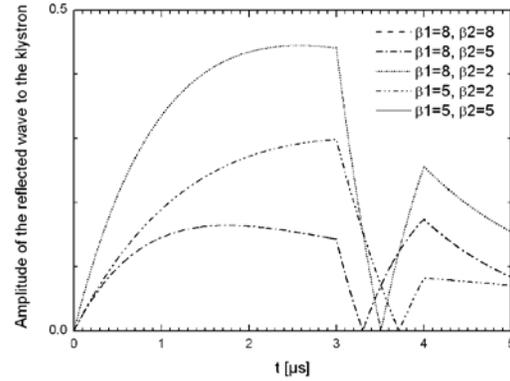

Fig. 16: Amplitude of the reflected wave to the klystron

## 5. Conclusions

Theoretical studies on the conventional SLED type of rf pulse compressor have been done. With the derived generalized formulae, the error and jitter effects on the SLED performance for the BEPCII linac have been studied systematically. PSK switching phase jitter, cooling water temperature variation and frequency unbalance error affect both the amplitude and phase of the outgoing wave. In the meantime, frequency unbalance will produce power reflection to the klystron. PSK switching time jitter, $Q_0$ and β unbalance errors only affect the outgoing wave amplitude enhancement, but have not any effect on the phase. $Q_0$ and β unbalance errors also result in power reflection to the klystron; if they are accompanied by the frequency unbalance error, the outgoing wave phase deviation will be magnified or minified. Higher $Q_0$ and lower β will magnify the phase deviation, lower $Q_0$ and higher β will minify the phase deviation.

To understand the error and jitter effects on the SLED performance is very helpful for us to facilitate the mechanical fabrication, microwave tuning and on-line operation of the SLED. However, limited by the paper length, a perfect 3dB coupler is assumed in this paper, no related error effects are included in our analysis.